\documentclass[11pt]{amsart}
\usepackage{amscd}
\usepackage{amsmath}
\usepackage{graphicx}
\usepackage{amsfonts}
\usepackage{amssymb}
\begin{document}

\title[Criteria of   positivity for linear maps]
{Criteria of   positivity for linear maps constructed from
permutation pairs}

\author{Haili  Zhao, Jinchuan Hou}
\address[H. Zhao, J. Hou]{Faculty of Mathematics, Taiyuan University of Technology,
Taiyuan 030024, P. R. of China} \email{zhaohaili927@yahoo.com.cn;
jinchuanhou@yahoo.com.cn}

\thanks{This work is partially supported by  Natural
Science Foundation of China (11171249, 11271217)}

\begin{abstract}
In this paper, we show that a $D$-type map $\Phi_D:M_n\rightarrow
M_n$ with $D=(n-2)I_n+P_{\pi_1}+P_{\pi_2}$ induced by a pair
 $\{\pi_1,\pi_2\}$  of permutations of $(1,2,\ldots, n)$ is positive if
$\{\pi_1,\pi_2\}$ has property (C). The property (C) is
characterized for $\{\pi_1,\pi_2\}$, and an easy criterion is given
for the case that $\pi_1=\pi^p$ and $\pi_2=\pi^q$, where $\pi$ is
the permutation defined by $\pi(i)=i+1$ mod $n$ and $1\leq p<q\leq
n$.

\end{abstract}
\maketitle

\section{Introduction}

Denote by   $M_{m,n}$ the set of all $m\times n$ complex matrices,
and write $M_{n,n}=M_n$ and $ M_n^+$ the set of all positive
semi-definite matrices in $M_n$. A linear map $L: M_n\rightarrow
M_n$ is positive if $L(M_n^+)\subseteq M_n^+$.  The study of
positive maps have been the central theme for many pure and applied
topics. For example, see \cite{CJW,Hor,HLPQS,MM,KK}. In particular,
the study is pertinent in quantum information science research; see
\cite{AU,ABCL,GW,HJ,LS,LP,NC}. So, the study of positive maps is
significant.

 Suppose $\Phi_D:{M_n}\rightarrow {M_n}$ is a linear map of the
form
$$(a_{ij})\longmapsto  {{\rm diag}(f_1,f_2,...,f_n)-(a_{ij})}
  \eqno(1.1)$$  with $(f_1,f_2,...,f_n)=(a_{11},a_{22},...,a_{nn})D
$
 for an $n\times n$ nonnegative matrix $D=(d_{ij})$ (i.e.,
 $d_{ij}\geq0$ for all $i,j$).
The map $\Phi_D$ of the form Eq.(1.1) defined by a nonnegative
matrix $D$ is called a $D$-type map.    The question of when a
$D$-type map is positive was studied intensively by many authors and
applied in quantum information theory to detect entangled states and
construct entanglement witnesses (ref., for instance,
\cite{HLPQS,QH} and the references therein).

A very interesting class of $D$-type maps is the class of maps
constructed from permutations.

Assume that $\pi$ is a permutation of $(1,2,\ldots,n)$. Recall that
the permutation matrix $P_{\pi}=(p_{ij})$ of $\pi$ is a $n\times n$
matrix determined by
$$p_{ij}=\begin{cases}
1 \ \ \ \ \ \ {\rm if}\ \  i=\pi(j)\ ({\rm mod}\ n),\\
0 \ \ \ \ \ \ {\rm if}\ \  i\neq{\pi(j)}\ ({\rm mod}\ n).
\end{cases}$$
Recall also that a subset  $\{i_1, \dots,
i_l\}\subseteq\{1,2,\ldots,n\}$ is an $l$-cycle of the permutation
$\pi$  if $\pi(i_j) = i_{j+1}$ for $j = 1, \dots, l-1$ and $\pi(i_l)
= i_1$. Note that every permutation $\pi$ of $(1,\dots, n)$ has a
disjoint cycle decomposition $\pi=(\pi_1)(\pi_2) \cdots (\pi_r)$,
that is, there exists a set $\{F_s\}_{s=1}^r$ of disjoint cycles of
$\pi$ with $\cup_{s=1}^r F_s=\{1,2,\ldots , n\}$ such that
$\pi_s=\pi|_{F_s}$ and $\pi(i)=\pi_s(i)$ whenever $i\in F_s$.

The positivity of $D$-type maps generated by one permutation were
discussed in \cite{HLPQS} recently. Let $\pi$ be a permutation of
$(1,2,\ldots ,n)$ with disjoint cycle decomposition $(\pi_1) \cdots
(\pi_r)$ such that the maximum length of $\pi_h$s is equal to $l>1$
and $P_{\pi}=(\delta_{i\pi(j)})$ is the permutation matrix
associated with $\pi$. For $t\geq 0$, let $\Phi_{t,\pi}: M_n
\rightarrow M_n$ be the $D$-type map of the form in Eq.(1.1) with $D
= (n-t)I_n + tP_{\pi}$. It is shown in \cite{HLPQS} that
$\Phi_{t,\pi}$ is positive if and only if $0\leq t\leq \frac{n}{l}$.
Particularly, $\Phi_\pi=\Phi_{1,\pi}$ is always positive.

Motivated by the above result, let us consider the $D$-type maps
constructed from several permutations. Generally speaking,
$\Phi_{D_{\pi_1,\ldots,\pi_k}}$ is not positive if
$$D_{\pi_1,\ldots,\pi_k}=(n-k)I_n+P_{\pi_1}+P_{\pi_2}+\cdots
+P_{\pi_k} \eqno(1.2)$$ with $1<k\leq n-1$. For example, take $\pi$
defined by $\pi(i)=i+1$ mod $n$ and let $D=(n-2)I_n+2P_{\pi}$,; then
$\Phi_D$ is not positive by the result in \cite{HLPQS} mentioned
above. However there do exist $D$-type positive linear maps of the
form $\Phi_{D_{\pi_1,\ldots,\pi_k}}$ with $k>1$. For the permutation
$\pi$ defined by $\pi(i)=i+1$ mod $n$, and for any $1\leq k\leq
n-1$, it is known that $\Phi_D : M_n\rightarrow M_n$ is positive,
where $D=(n-k)I_n+P_\pi+P_{\pi^2}+\cdots+P_{\pi^k}$ (ref.
\cite{SY}).

So it is an natural and interesting topic to study the $D$-type
linear maps   $\Phi_D$ with $D={D_{\pi_1,\ldots,\pi_k}}$ of the form
Eq.(1.2) and establish positivity criteria for them.

The purpose of this paper is to start the discussion of this topic.
We introduce a concept of property (C) for a pair of permutations
(Definition 3.2), and show that, if $\{\pi_1, \pi_2\}$ has the
property (C), then the $D$-type map $\Phi_{D_{\pi_1, \pi_2}}$ is
positive (Theorem 3.3 and Theorem 3.4). To do this, we develop some
inequalities (Lemma 2.1, Lemma 2.2), which are interesting
themselves. Furthermore, we give a necessary and sufficient
condition for two permutations to have property (C) (Proposition
3.5). This results then applied to give some criteria of positivity
of $D$-type maps induced by pair of permutations and to construct
some new positive $D$-type maps, especially from pair of
permutations $\{\pi^p,\pi^q\}$, where $\pi$ is the cyclic
permutation defined by $\pi(i)=i+1$ mod $n$. Some examples are also
presented to illustrate how to use the criteria.

\section{Preliminary inequalities}

In this section, we first establish two cyclic like inequalities,
which are needed to establish the criteria   for a $D$-type linear
maps induced from a pair of permutations to be positive.

 \textbf{Lemma 2.1.} {\it Let $s,M$ be   positive numbers and $f(u_1,u_2,...,u_m)$ be a function in $m$-variable
defined by
$$f(u_1,u_2,...,u_m)=\frac 1{s+u_1}+\frac
1{s+u_2}+...+\frac1{s+u_m}$$ on the region $u_i>0$ with
$u_1u_2...u_m=M^m$, $i=1,2,...,m$. Then we have}

(1)  {\it  $f$ has extremum values $
\frac{rs^{\frac{m}{2r-m}}+(m-r)M^{\frac{m}{2r-m}}}{s(s^{\frac{m}{2r-m}}+M^{\frac{m}{2r-m}})}$
with $\frac{m}{2}< r\leq m$ at the points that $r$ of $u_i$s are
$(\frac{M^m}{s^{2m-2r}})^{\frac{1}{2r-m}} $ and others are
$(\frac{s^{2r}}{M^m})^{\frac{1}{2r-m}} $; }

(2) {\it $f$ may also achieve the extremum $\frac{m}{s+M}$    when
$m$ is even, at points  $\frac{m}{2}$ of $u_i$s are $u$ and others
are $\frac{s^2}{u}$, in this case we must have $s=M$;}

(3) $\sup f(u_1,u_2,...,u_m)=\max\{\frac{m-1}{s}, \frac{m}{s+M}\}.$

\if

(4) {\it  if $ M\leq\frac{s}{m-1}$, then $\max
f(u_1,u_2,...,u_m)=f(M,M,\ldots,M)=\frac{m}{s+M}$;}

(5) {\it  if $ M>\frac{s}{m-1}$, then $\sup f(u_1,u_2,...,u_m)=
\frac{m-1}{s}$ . \fi

 {\bf Proof.} The case $m=1$ is obvious. For the case that $m= 2$,
 we have $u_2=\frac{M^2}{u_1}$ and $f(u_1,u_2)=\frac 1{s+u_1}+\frac
1{s+\frac{M^2}{u_1}}=\frac 1{s+u_1}+\frac {u_1}{su_1+ M^2 }
=g(u_1)$. Letting $g^\prime
(u_1)=\frac{M^2}{(su_1+M^2)^2}-\frac{1}{(s+u_1)^2}=0$, we get
$(M^2-s^2)(u_1^2-M^2)=0$. If $M\not=s$, we must have $u_1=M$ and $g$
has an extremum at $u_1=M$ with $  g(M)=\frac{2}{s+M}$; if $s=M$,
then $g(u_1)=\frac{1}{s+u_1}+\frac{u_1}{s(s+u_1)}=\frac{1}{s}$. Also
note that $\lim_{u_1\rightarrow 0} g(u_1)=\frac{1}{s}$. So, $\max
f(u_1,u_2)\leq\max\{\frac{1}{s},\frac{2}{s+M}\}$.

Assume that $m\geq 3$. $f$ approximates its supremum at the
singular boundary of its domain or at some of its extremum values.
It is obvious that if $(u_{1j},u_{2j},\ldots, u_{mj})$ converges
to the boundary of the domain $\{ \Pi _{i=1}^m u_i=M^m\}$, then
some $u_i$ tend to 0 and some  tend to $\infty$. Say $k$ of them
tend to 0 and $m-k-r$ of them tend to $\infty$. Thus there are $r$
nonnegative numbers $t_h$ so that
$\overline{\lim}_{j\rightarrow\infty} f(u_{1j},u_{2j},\ldots,
u_{mj})=\frac{k}{s}+\sum_{h=1}^r\frac{1}{s+t_h}\leq
\frac{k+r}{s}\leq \frac{m-1}{s}$.

Next let us consider the extremum values of $f$.  Let
$\varphi(u_1,u_2,...,u_m)=u_1u_2...u_m-M^m,$ and
 $$\begin{array}{rl}
 L(u_1,u_2,...,u_m,\lambda)=&f(u_1,u_2,...,u_m)+\lambda\varphi(u_1,u_2,...,u_m)\\
=& \sum_{i=1}^m \frac
 {1}{s+u_i}+\lambda(u_1u_2...u_m-M^m).\end{array}$$
 By the method of Lagrange multipliers, we have the system
 $$L_{u_i}^\prime=\frac{-1}{(s+u_i)^2}+\lambda{u_1u_2...u_{i-1}u_{i+1}...u_m}=\frac{-1}{(s+u_i)^2}+\frac{\lambda M^m}{u_i}=0,\eqno(2.1)$$
 $i=1,2\ldots,m$.
\if Let $$L_{u_i}^\prime=\frac{-1}{(s+u_i)^2}+\frac{\lambda
 M^k}{u_i}=0\eqno(2.1)$$
 for $i=1,2,...,k$. \fi
 Solving this system gives $${\lambda M^m}=\frac{u_i}{(s+u_i)^2}\eqno(2.2)$$ for
 $i=1,2,...,m.$ Thus, for any $i,j$ with $i\not=j$,  $\frac{u_i}{(s+u_i)^2}={\lambda M^m}=\frac{u_j}{(s+u_j)^2}$
which yields  $$u_i[s+u_j]^2=u_j[s+u_i]^2.$$ It follows that
                 $${s^2}{u_i}+2{u_i}{u_j}{s}+{u_i}{u_j}^2={s^2}{u_j}+{2u_iu_js}+{u_j}{u_i^2},$$
                             $${s^2}(u_i-u_j)+u_iu_j(u_j-u_i)=0,$$
                                      $$(u_i-u_j)[s^2-u_iu_j]=0.$$
 Hence we get                                     $${u_i=u_j} \ \ \ {\rm or} \ \ \ {u_iu_j=s^2}.\eqno(2.3)$$
Eq.(2.3) implies that, there exists a positive integer
$\frac{m}{2}\leq r\leq m$ such that $r$ of $ u_1,u_2,\ldots, u_m $
are the same, denoted by $u$, and other $m-r$ of them are the
same, denoted by $v$, with $uv=s^2$. Note that
$$u^{(2r-m)}s^{2(m-r)}=u^rv^{m-r} =u_1u_2\cdots u_m=M^m. \eqno(2.4)$$

If $2r=m$ (in this case  $m$ is even), then we get $s^m=M^m$ and
hence $ s=M$. Consequently, $f$ has a possible extremum value
$$\frac{m}{2(s+u)}+\frac{m}{2(s+\frac{s^2}{u})}=\frac{m}{2(s+u)}+\frac{mu}{2s(s+u)}=\frac{m}{2s}=\frac{m}{s+M}$$
at each point $(u_1,u_2,\ldots,u_m)$ satisfying that $\frac{m}{2}$
of $u_i$s are $u$ for some $u>0$ and other $\frac{m}{2}$ of them are
$ \frac{s^2}{u}$.

If $m<2r\leq 2m$, we get by Eq.(2.4) that
$$u=\left(\frac{M^m}{s^{2(m-r)}}\right)^{\frac{1}{2r-m}}\quad {\rm
and} \quad v=\left(\frac{s^{2r}}{M^{m}}\right)^{\frac{1}{2r-m}},
\eqno(2.5)$$ and hence   $f$ has an extremum value
$$\begin{array}{rl} \frac{r}{s+u}+\frac{m-r}{s+v}= &\frac{r}{s+\left(\frac{M^m}{s^{2(m-r)}}\right)^{\frac{1}{2r-m}}}+
\frac{m-r}{s+\left(\frac{s^{2r}}{M^{m}}\right)^{\frac{1}{2r-m}}}
\\ =& \frac{rs^{\frac{2m-2r}{2r-m}}+\frac{m-r}{s}M^{\frac{m}{2r-m}}}{s^{\frac{m}{2r-m}}+M^{\frac{m}{2r-m}}} =
\frac{rs^{\frac{m}{2r-m}}+(m-r)M^{\frac{m}{2r-m}}}{s(s^{\frac{m}{2r-m}}+M^{\frac{m}{2r-m}})}
\end{array} \eqno(2.6)$$
at those points $(u_1,u_2,\ldots, u_m)$ that $r$ of $u_i$s equal
$\left(\frac{M^m}{s^{2(m-r)}}\right)^{\frac{1}{2r-m}}$ and other
$m-r$ of $u_i$s equal
$\left(\frac{s^{2r}}{M^{m}}\right)^{\frac{1}{2r-m}}$.

Note that $1<\frac{m}{2}\leq r\leq m$ and
$$\frac{m-1}{s}-\frac{rs^{\frac{m}{2r-m}}+(m-r)M^{\frac{m}{2r-m}}}{s(s^{\frac{m}{2r-m}}+M^{\frac{m}{2r-m}})}
=\frac{(m-1-r)s^{\frac{m}{2r-m}}+(r-1)M^{\frac{m}{2r-m}}}{s(s^{\frac{m}{2r-m}}+M^{\frac{m}{2r-m}})}.$$
 So, if $r\leq m-1$, then we always have
$\frac{m-1}{s}\geq\frac{rs^{\frac{m}{2r-m}}+(m-r)M^{\frac{m}{2r-m}}}{s(s^{\frac{m}{2r-m}}+M^{\frac{m}{2r-m}})}$.
If $r=m$, then
$\frac{rs^{\frac{m}{2r-m}}+(m-r)M^{\frac{m}{2r-m}}}{s(s^{\frac{m}{2r-m}}+M^{\frac{m}{2r-m}})}=\frac{m}{s+M}$.
These, together with the cases that $m$ is even and $r=\frac{m}{2}$
entails that $\sup f(u_1,u_2,\ldots, u_m)\leq
\max\{\frac{m-1}{s},\frac{m}{s+M}\}$, completeing the proof.
\hfill$\Box$

\if We claim that $\frac{m}{s+M}$ is the largest one in the extremum
values whenever $s\geq M$. In fact, for $\frac{m}{2}<k\leq m-1$, as
$\frac{m}{2k-m}>1$, we have by Eq.(2.5)
$$\frac{ks^{\frac{m}{2k-m}}+(m-k)M^{\frac{m}{2k-m}}}{s(s^{\frac{m}{2k-m}}+M^{\frac{m}{2k-m}})}
<\frac{m}{s(1+(\frac{M}{s})^{\frac{m}{2k-m}})}<\frac{m}{s(1+\frac{M}{s})}
=\frac{m}{s+M}.
$$
So (2)is true.  Further more, observe that $f(M,M,\ldots,
M)=\frac{m}{s+M}$, $\frac{m-1}{s}\leq \frac{m}{s+M}$ if and only if
$M\leq \frac{s}{m-1}$. Hence the last two assertions are also
true.\fi

The following lemma is crucial for our purpose. Though we only
need the special case of $k=2$ in this paper, we present the
inequality for any $k\geq 1$ because it may be useful to
discussing $D$-type maps induced by $k$ permutations.

\textbf{Lemma 2.2.} {\it Let $s$ be a positive number, $n, k$ be
positive integers with $s>k$. Then for any $nk$  positive real
numbers $\{x_{hi} ,h=1,2,...,k; i=1,2,...,n,\}$ satisfying
$x_{h1}x_{h2}...x_{hn}=1$ for each $h$ with $1\leq h\leq k$, we have
$$\begin{array}{rl} f(x_{11},\ldots, x_{1n},x_{21},\ldots, x_{k1},\ldots, x_{kn})=&\sum_{i=1}^n
\frac{1}{s-k+x_{1i}+x_{2i}+...+x_{ki}}\\ \leq
&\max\{\frac{n-1}{s-k}, \frac{n}{s}\}.\end{array}\eqno(2.7)$$
Moreover, the  extremum values of $f$ are}
$$\begin{array}{ll} \delta_r=\frac{r(s-k)^{\frac{n}{2r-n}}+(n-r)k^{\frac{n}{2r-n}}}{(s-k)((s-k)^{\frac{n}{2r-n}}+k^{\frac{n}{2r-n}})},
& [\frac{n}{2}]+1\leq r< n; \\
 \delta_{\frac{n}{2}}=\frac{n}{s} & \mbox{\rm if } n\ \mbox{\rm is
even},
\end{array} \eqno(2.8)
$$
{\it where $[t]$ stands for the integer part of real number $t$.}

 \if $f$ achieves the extremum $\delta_n $ at
$x_{11}=x_{21}=\ldots=x_{k,n}=1$; $f$ achieves $\delta_r$ with
$\frac{n}{2}<r\leq n-1$ at those points that
$x_{1i}=x_{2i}=\ldots=x_{ki}=v_i$ for all $i=1,2,\ldots, n$ and $r$
of $v_i$s equal $(\frac{k}{s-k})^{\frac{2n-2r}{2r-n}}$ and other
$n-r$ $v_i$s equal $(\frac{s-k}{k})^{\frac{2r}{2r-n}}$; and, if $n$
is even, $f$ achieves the extremum $\delta_{\frac{n}{2}}$ at those
points that $\frac{n}{2}$ of $v_i$s are the same, denoted by $v$,
and other $\frac{n}{2}$ $v_i$s are equal to $\frac{(n-k)^2}{v}$, in
this case we must have $k=\frac{n}{2}=r$.\fi

 {\bf Proof.}  The question is reduced to find the supremum  of the
$kn$-variable function
 $$f(x_{11},...,x_{1n},x_{21},...,x_{2n},...,x_{k1},...,x_{kn})=\sum_{i=1}^n\frac{1}{s-k+x_{1i}+x_{2i}+...+x_{ki}}$$
 on the region $x_{hi}>0$ with
$x_{h1}x_{h2}...x_{hn}=1$, $h=1,2,\ldots, k$.   Considering the
behavior of $f$ near the boundary of its domain, it is obvious that
the upper bound of $f$ near the boundary is $\frac{n-1}{s-k}$.

To find the extremum values of $f$, let
$$\begin{array}{rl}
&L(x_{11},...x_{1n},x_{21},...,x_{2n},...,x_{k1},...,x_{kn},\lambda_1,...\lambda_k)\\
=&f(x_{11},...x_{1n},x_{21},...,x_{2n},...,x_{k1},...,x_{kn})+\sum_{h=1}^k\lambda_h\varphi_h,\end{array}$$
where $\varphi_h=x_{h1}x_{h2}...x_{hn}-1$, $h=1,2,...k.$ By the
method of Lagrange multipliers, we have the system
$$\begin{cases}L_{x_{11}}^\prime=\frac{-1}{[(s-k)+x_{11}+x_{21}+...+x_{k1}]^2}+\frac{\lambda_1}{x_{11}}=0,\\
\ \vdots\\
L_{x_{k1}}^\prime=\frac{-1}{[(s-k)+x_{11}+x_{21}+...+x_{k1}]^2}+\frac{\lambda_k}{x_{k1}}=0,\\
\\
L_{x_{12}}^\prime=\frac{-1}{[(s-k)+x_{12}+x_{22}+...+x_{k2}]^2}+\frac{\lambda_1}{x_{12}}=0,\\
\ \vdots\\
L_{x_{k2}}^\prime=\frac{-1}{[(s-k)+x_{12}+x_{22}+...+x_{k2}]^2}+\frac{\lambda_k}{x_{k2}}=0,\\
\ \vdots\\
L_{x_{1n}}^\prime=\frac{-1}{[(s-k)+x_{1n}+x_{2n}+...+x_{kn}]^2}+\frac{\lambda_1}{x_{1n}}=0,\\
\ \vdots\\
L_{x_{kn}}^\prime=\frac{-1}{[(s-k)+x_{1n}+x_{2n}+...+x_{kn}]^2}+\frac{\lambda_k}{x_{kn}}=0.\end{cases}\eqno(2.7)$$
Solving this system, one obtains that
$$
\begin{cases}
\frac{\lambda_1}{x_{11}}={\frac{\lambda_2}{x_{21}}}=\ldots={\frac{\lambda_k}{x_{k1}}},\\
{\frac{\lambda_1}{x_{12}}}={\frac{\lambda_2}{x_{22}}}=\ldots={\frac{\lambda_k}{x_{k2}}},\\
\vdots\\
{\frac{\lambda_1}{x_{1n}}}={\frac{\lambda_2}{x_{2n}}}=\ldots={\frac{\lambda_k}{x_{kn}}},
\end{cases}\eqno(2.9) $$
which implies that
$$\frac{{\lambda_1}^n}{{x_{11}}{x_{12}}\ldots{x_{1n}}}=\frac{{\lambda_2}^n}{{x_{21}}{x_{22}}\ldots{x_{2n}}}
=\ldots=\frac{{\lambda_k}^n}{{x_{k1}}{x_{k2}}\ldots{x_{kn}}}.\eqno(2.10)$$\\
Now since ${x_{h1}}{x_{h2}}\ldots{x_{hn}}=1$, for each
$h=1,2,...,k$, we get
$${{\lambda_1}^n}={{\lambda_2}^n}=\ldots={{\lambda_k}^n}.\eqno(2.11)$$\\
Furthermore, by Eq.(2.8) we see that
$${\lambda_h}={\frac{x_{h1}}{{[(s-k)+({x_{11}}+{x_{21}}+\ldots+{x_{k1}})]}^2}}=\ldots={\frac{x_{hn}}{{[(s-k)+({x_{1n}}+{x_{2n}}+\ldots+{x_{kn}})]}^2}}>0$$
for each $h=1,2,...,k.$ Therefore, we must have
$${\lambda_1}={\lambda_2}=\ldots={\lambda_k}>0,$$
which forces, by the Eq.(2.9), that
$$\begin{cases}
{x_{11}}=x_{21}=\ldots=x_{k1},\\
{x_{12}}=x_{22}=\ldots=x_{k2},\\
\ \vdots\\
{x_{1n}}=x_{2n}=\ldots=x_{kn}.\\
\end{cases}\eqno(2.11)$$\\
 Let $v_i=x_{1i}$, $i=1,2,...,n.$ Then,
 the question reduces to find the supremum of the
function in $ n$-variable
$$g(v_1,v_2,...,v_n)=\frac 1{s-k+kv_1}+\frac
1{s-k+kv_2}+...+\frac1{s-k+kv_n}$$ on the region
 $v_i>0$ and $v_1v_2\cdots
 v_n=1$. Applying Lemma 2.1 with $m=n$, $s$ replaced by $s-k$, $M=k$ and $u_i=kv_i$, $i=1,2, \ldots n$,  we obtain that
$$\sup g(v_1,v_2,...,v_n)=\max\{\frac{n-1}{s-k}, \frac{n}{s-k+k}\}=\max\{\frac{n-1}{s-k},
 \frac{n}{s}\}.$$
 Hence $\sup
f=\max\{\frac{n-1}{s-k},
 \frac{n}{s}\}$.
Moreover, the extremum values of $f$ are $\{\delta_r: n\leq 2r\leq
2n\}$, where
$$
\delta _n=\frac{n}{s},\
\delta_r=\frac{r(s-k)^{\frac{n}{2r-n}}+(n-r)k^{\frac{n}{2r-n}}}{(s-k)((s-k)^{\frac{n}{2r-n}}+k^{\frac{n}{2r-n}})}\quad{\rm
for}\ [\frac{n}{2}]+1\leq r\leq n-1,
$$
 and $$\delta_{\frac{n}{2}}=\frac{n}{s} \quad{\rm if}\ n \ {\rm is\ even}.$$
It is easily checked by Lemma 2.1 that $f$ achieves the extremum
$\delta_n $ at $x_{11}=x_{21}=\ldots=x_{k,n}=1$; $f$ achieves
$\delta_r$ with $\frac{n}{2}<r\leq n-1$ at those points that $r$ of
$v_i$s equal $(\frac{k}{s-k})^{\frac{2n-2r}{2r-n}}$ and other $n-r$
of $v_i$s equal $(\frac{s-k}{k})^{\frac{2r}{2r-n}}$; and, if $n$ is
even, $f$ achieves the extremum $\delta_{\frac{n}{2}}$ at those
points that $\frac{n}{2}$ of $v_i$s are the same, denoted by $v$,
and other $\frac{n}{2}$ $v_i$s are equal to $\frac{(s-k)^2}{v}$, in
this case we must have $k=\frac{n}{2}=r$. \hfill$\Box$

\section{Criteria of positivity of $D$-type maps induced by two permutations}

 In \cite{HLPQS}, the authors established a criterion of positivity for $D$-type maps constructed from one permutation, and proved
that $\Phi_D:M_n\rightarrow M_n$ is positive if $D=(n-1)I_n+P_\pi$,
where   $ \pi $ is any permutation of $\{1,2,...,n\}$. For those
constructed from more than one permutations, it is known that, the
$D$-type map
  $\Phi_D$ with
$D=(n-k)I_n+P_{\pi_0}+P_{{\pi_0}^2}+...+P_{{\pi_0}^k}$ is positive,
where $ {\pi_0}(i)= i-1 $ mod $n$ \cite{ABCL} or $\pi_0(i)=i+1$ mod
$n$ \cite{SY}. In this section we    give a criterion of positivity
of $D$-type linear maps constructed from pairs of general
permutations.

The following lemma comes from \cite{HLPQS}.

\textbf{Lemma  3.1.} {\it Suppose $\Phi_D:{M_n}\rightarrow {M_n}$ is
a $D$-type linear map of the form
$$(a_{ij})\longmapsto  {{\rm diag}(f_1,f_2,...,f_n)-(a_{ij})}
  \eqno(3.1)$$  with $(f_1,f_2,...,f_n)=(a_{11},a_{22},...,a_{nn})D
$
 for an $n\times n$ nonnegative matrix $D=(d_{ij})$ (i.e.,
 $d_{ij}\geq0$ for all $i,j$).
 Then, $\Phi_D$ is positive if and only if, for any unit vector
 $u=(u_1,u_2,...,u_n)^t\in {\mathbb C}^n$, we have $f_j(u)=\sum_{i=1}^n
 d_{ij}|u_i|^2\neq 0$ whenever $u_j\neq0$, and that
 $\sum_{u_j\neq0}{\frac{|u_j|^2}{f_j(u)}}\leq 1.$}

Before stating our main result, we introduce a conception of
property (C) for a pair  of permutations.

{\bf Definition 3.2.} {\it A pair  $\{\pi_1,\pi_2 \}$
 of permutations of $(1,2,\ldots, n)$  is said to have
 property (C) if,  for  any given $i\in\{1,2,\ldots, n\}$ and for any $j\not= i$, there exists $\pi_{h_j}(j)\in\{\pi_1(j),\pi_2(j)\}$
with $h_j\in\{1,2\}$ such that $\{\pi_{h_j}(j): j=1,2,\ldots,
i-1,i+1,\ldots, n\}=\{1,2,\ldots, i-1,i+1,\ldots, n\}$.}

  Let  $\pi$ be the permutation of $(1,2,\ldots, n)$
 defined   by $\pi(i)=i+1$ mod $n$. Then it is easily checked that, for any $1\leq p\leq n-1$, $\{\pi^p,\pi^{p+1}\}$ has Property (C).
 There are other kinds of examples. For instance,  $\{\pi,\pi^4\}$ has the property (C) if $n=5$ and $7$, but
 does  not have the  property (C) if $n=4$ and $6$.

The following is our main general result which gives a criterion of
positivity for $D$-type maps induced by pair of general
permutations.

\textbf{Theorem 3.3.}  {\it For any two permutations ${\pi_1}$ and
$\pi_2$ of
 $(1,2,...,n)$ with $n\geq 3$,
 let $\Phi_D : M_n\rightarrow M_n$ be the $D$-type  map of the form
 $$(a_{ij})\longmapsto  {{\rm diag}(f_1,f_2,...,f_n)-(a_{ij})},$$
where $(f_1,f_2,...,f_n)=(a_{11},a_{22},...,a_{nn})D $ and
$D=(n-2)I_n+{P_{\pi_1}}+{P_{\pi_2}}$  with $P_{\pi_h}$ the
permutation matrix of $\pi_h$, $h=1,2$. If $\{\pi_1,  \pi_2\}$ has
the property (C), then $\Phi_D $ is positive.}

{\bf Proof.} For any unit vector $x=(x_1,x_2,...,x_n)^t\in {\mathbb
C}^n$, we have
$$X=xx^*={(x_1,x_2,...,x_n)^t}(\bar{x_1},\bar{x_2},...\bar{x_n})=\left(
\begin{array}{cccc}
 |x_{1}|^2&{x_1\bar{x_2}}&\cdots&{x_1\bar{x_n}}\\
{x_2\bar{x_1}}&|x_{1}|^2&\cdots&{x_2\bar{x_n}}\\
\vdots&\vdots&\ddots&\vdots\\
{x_n\bar{x_1}}&{x_n\bar{x_2}}&\ldots&|x_n|^2 \end{array} \right).
$$
As $D=(d_{ij})=(n-k)I_n+{P_{\pi_1}}+{P_{\pi_2}}$, we get
$$f_j(x)=\sum_{i=1}^nd_{ij}|x_i|^2\\
=(n-2)|x_j|^2+|x_{\pi_1(j)}|^2+|x_{\pi_2(j)}|^2.$$ Then by Lemma
3.1,  $\Phi_D$ is positive if the following inequality
$$\sum_{i=1}^n\frac{|x_i|^2}{f_i(x)}=
\sum_{i=1}^n{\frac{|x_i|^2}{(n-2)|x_i|^2+|x_{\pi_1(i)}|^2+|x_{\pi_2(i)}|^2}}\leq
1\eqno(3.2)$$ holds for any unit vector
$x=(x_1,x_2,\ldots,x_n)^t\in{\mathbb C}^{(n)}$.

{\bf Case 1.} $n\geq 4$.

Let us consider first the case that $n\geq 4$. Assume that all
$x_i\neq0$ and let $u_{hi}={\frac{|x_{\pi_h(i)}|^2}{|x_i|^2}}$, $
i=1,2,...,n$; $h=1,2.$ Then, $u_{h1}u_{h2}\cdots u_{hn}=1$ for each
$h=1,2$, and thus, by Lemma 2.2, we have
$$\begin{array}{rl} &f(x_1,x_2,\ldots, x_n)=\sum_{i=1}^n{\frac{|x_i|^2}{(n-2)|x_i|^2+|x_{\pi_1(i)}|^2+|x_{\pi_2(i)}|^2}} \\
 =&g(u_{11},u_{12},\ldots, u_{1n},u_{21},u_{22},\ldots, u_{2n})=\sum_{i=1}^{n}{\frac{1}{n-2+u_{1i}+u_{2i}}}\\ \leq & \max\{\frac{n-1}{n-2},1\}. \end{array}$$
and, as a restriction of $g$,   the maximum of  all extremum
 values of $f$ is bounded by the maximum of all extremum values of
 $g$. Thus,  by Lemma 2.2, we have
 $$\max \{\mbox{\rm extreme values of } f\}\leq\max\{1,
 \delta_r: r=[\frac{n}{2}]+1,  [\frac{n}{2}]+1,\ldots, n-1\},
 $$
where
$$\delta_r=\frac{r(n-2)^{\frac{n}{2r-n}}+(n-r)2^{\frac{n}{2r-n}}}{(n-2)((n-2)^{\frac{n}{2r-n}}+2^{\frac{n}{2r-n}})}
$$ and $[t]$ is the integer part of $t$.
If $[\frac{n}{2}]+1\leq r\leq n-2$, then
$$\begin{array}{rl}\delta_r= &
\frac{r(n-2)^{\frac{n}{2r-n}}+(n-r)2^{\frac{n}{2r-n}}}{(n-2)((n-2)^{\frac{n}{2r-n}}+2^{\frac{n}{2r-n}})}\\
=&
\frac{(2r-n)(n-2)^{\frac{n}{2r-n}}}{(n-2)((n-2)^{\frac{n}{2r-n}}+2^{\frac{n}{2r-n}})}+\frac{n-r}{n-2}\leq\frac{r}{n-2}\leq
1.\end{array}
$$
We claim that we also have $\delta_{n-1}\leq 1$. In fact,
$$\delta_{n-1}=\frac{(n-1)(n-2)^{\frac{n}{2r-n}}+
2^{\frac{n}{2r-n}}}{(n-2)((n-2)^{\frac{n}{2r-n}}+2^{\frac{n}{2r-n}})}=\frac{(n-2)^{\frac{n}{2r-n}}}{(n-2)^{\frac{n}{2r-n}}+2^{\frac{n}{2r-n}}}+\frac{1}{n-2}\leq
1
$$
if and only if
$$\psi(n)=\frac{n-2}{(n-3)^{\frac{n-2}{n}}}\leq 2. \eqno(3.3)$$
Let $\varphi (t)=\ln
(\frac{t-2}{(t-3)^{\frac{t-2}{t}}})=\ln(t-2)-\frac{t-2}{t}\ln
(t-3)$ for $t\geq 4$. Then $$\varphi^\prime
(t)=\frac{t^2-4t-(2t^2-10t+12)\ln(t-3)}{t^2(t-2)(t-3)}.$$ If
$t\geq 9$, then
$$t^2-4t-(2t^2-10t+12)\ln(t-3)<-(t^2-10t+12)\ln(t-3)<0$$ and hence
$\varphi(t)$ is decreasing on $[9,\infty)$. It follows that
$\psi(n)=\frac{n-2}{(n-3)^{\frac{n-2}{n}}}$ is decreasing for $
n\geq 9$. Since
$$\psi(4)=2, \ \psi(5)\approx 1.9793,\  \psi(6)\approx 1.9230,\ \psi(7)\approx 1.8575,\ \psi(8)\approx 1.7944,\ \psi(9)\approx 1.7373,
$$
we see that the inequality in Eq.(3.3) is true for all $n\geq 4$ and
hence $\delta_{n-1}\leq 1$ holds for all $n\geq 4$, as desired.

 Therefore, for any $n\geq 4$, 1 is the
maximum extremum value of $f(x_1,x_2,\ldots, x_n)$ achieving at
 $f(\frac{1}{\sqrt{n}}, \frac{1}{\sqrt{n}},\ldots,\frac{1}{\sqrt{n}})$.

\if false Now consider the case of $n=4$. Thus $1\leq s\leq 3$.
Denote $r_l=\#{\mathcal F}_l$, $l=1,\ldots, s$.

If one of  $\pi_1(4)$ and $\pi_2(4)$ is  $4$, say $\pi_1(4)=4$,
then $y_{\pi_2(4)}=y_4$, and we have
$$ \begin{array}{rl} & \sum_{i=1}^4\frac{y_i}{2y_i+y_{\pi_1(i)}+y_{\pi_2(i)}}=\sum_{l=1}^s\sum_{i\in{\mathcal
F}_l}
\frac{y_i}{2y_i+y_{\pi_1(i)}+y_{\pi_2(i)}}+\frac{y_4}{2y_4+y_{\pi_1(4)}+y_{\pi_2(4)}}\\
=& \sum_{l=1}^s\sum_{i\in{\mathcal F}_l}
\frac{z_l}{2z_l+2z_l}+\frac{y_4}{2y_n+2y_4}=\sum_{l=1}^s\frac{r_l}{4}+\frac{1}{4}=\frac{3}{4}+\frac{1}{4}=1.
\end{array}$$

Assume both of $\pi_1(4)$ and $\pi_2(4)$ are not $4$. We consider
tree cases that may occur.

Case 1$^\circ$. $s=1$. Then $\sigma$ is cyclic and hence
$y_1=y_2=y_3$. This forces  that $y_4=y_1=y_2=y_3$ and hence $
\sum_{i=1}^4\frac{y_i}{2y_i+y_{\pi_1(i)}+y_{\pi_2(i)}}=1$.

Case 2$^\circ$. $s=3$. Then $\sigma (i)=i$ for $i=1,2,3$. Say
$\pi_1(1)=1$. Thus $\pi_1$ has four possible choices:
$$ (1,2,4,3), (1,3,4,2), (1,4,2,3), (1,4,3,2).$$
If $\pi_2(1)=1$, since $\{\pi_1,\pi_2\}$ has the property (C), the
possible choice of $\{\pi_1,\pi_2\}$ is only
$$\{(1,2,4,3), (1,4,3,2)\}.
$$
Thus we must  have $y_1=y_2=y_3=y_4$.

If $\pi_2(1)=2$, then $\pi_2$ has four choices:
$$(2,1,4,3), (2,3,4,1), (2,4,1,3), (2,4,3,1).$$ Thus
the suitable choice of $\{\pi_1,\pi_2\}$  is
$$\{1,2,4,3),(2,4,3,1)\},
$$ and
similarly  $\pi_2(1)=3$ implies that $\{\pi_1,\pi_2\}=\{(1,4,3,2),
(3,2,4,1)\}$. Thus $\pi_2(1)=2$ or $3$ will force
$y_1=y_2=y_3=y_4$.

The situation of $\pi_2(1)=4$ is slightly complicated. In this case,
$\{\pi_1,\pi_2\}$ can be any one of six pairs:
$$\begin{array}{lll}\{(1,2,4,3), (4,1,3,2)\},& \{(1,3,4,2),
(4,2,3,1)\},& \{(1,4,2,3),(4,2,3,1)\}, \\
 \{(1,4,3,2),(4,2,1,3)\} , &\{(1,2,4,3),(4,2,3,1)\} , &\{(1,4,3,2),(4,2,3,1)\}.\end{array}$$ The former four cases obviously imply
$y_1=y_2=y_3=y_4$ and hence
$\sum_{i=1}^4\frac{y_i}{2y_i+y_{\pi_1(i)}+y_{\pi_2(i)}}=1$.  The
last two subcases also imply the sum is equal to 1. For instance,
say  $\{\pi_1,\pi_2\}=\{(1,2,4,3),(4,2,3,1)\}$. Note that
$y_1=y_3=y_4$ and $\pi_1(2)=\pi_2(2)=2$, it is obvious that
$\sum_{i=1}^4\frac{y_i}{2y_i+y_{\pi_1(i)}+y_{\pi_2(i)}}=1$.

Case 3$^\circ$.  $s=2$. Then $\sigma $ has a fixed point. Without
loss of generality we assume that $\pi_1(1)=\sigma(1)=1$. Then
$\sigma(2)=3$ and $\sigma(3)=2$.

If $\pi_2(1)=1$, then $\{\pi_1,\pi_2\}= \{(1,3,4,2),(1,4,2,3)\}$;

If $\pi_2(1)=2$, then $\{\pi_1,\pi_2\}=\{(1,4,2,3),(2,3,4,1)\}$;

If $\pi_2(1)=3$, then $\{\pi_1,\pi_2\}=\{(1,3,4,2), (3,4,2,1)\}$;

If $\pi_2(1)=4$, then, since $\{(1,3,4,2),(4,3,2,1)\}$ and
$\{1,4,2,3), (4,3,2,1)\}$   do not have property (C), one sees that
$\{\pi_1,\pi_2\}$ is any one of $\{(1,2,4,3),(4,3,2,1)\}$,
$\{(1,3,4,2),(4,1,2,3)\}$, $\{1,4,2,3),(4,3,1,2)\}$ and
$\{(1,4,3,2),(4,3,2,1)\}$. Hence again we get $y_1=y_2=y_3=y_4$ and
$ \sum_{i=1}^4\frac{y_i}{2y_i+y_{\pi_1(i)}+y_{\pi_2(i)}}=1$. \fi

In the following we show that the supremum of  $f$   on the boundary
of its
 domain $|x_1|^2+|x_2|^2+\cdots+|x_n|^2=1$ is not greater than 1, either.
 Assume that $(x_1,x_2,\ldots, x_n)^t$ lies in the boundary of the region $|x_1|^2+|x_2|^2+\cdots+|x_n|^2=1$;
 then   at least one of $x_i$ is
 zero.  With no loss of generality, say $x_i\not=0$ for
$i=1,2,\ldots r$, but $x_{r+1}=\ldots=x_n=0$, where $1<r<n$.  Then
we have
$$\begin{array}{rl} &
\sum_{i=1}^{n}{\frac{|x_i|^2}{(n-2)|x_i|^2+|x_{\pi_1(i)}|^2+|x_{\pi_2(i)}|^2}}
=\sum_{i=1}^{r}{\frac{1}{(n-2)+\frac{|x_{\pi_1(i)}|^2}{|x_i|^2}+\frac{|x_{\pi_2(i)}|^2}{|x_i|^2}}}.
\end{array}$$
If $r=n-1$, by the assumption that $\{\pi_1,\pi_2\}$ has the
property (C), one can choose $\pi_{h_i}(i)\in\{\pi_1(i),\pi_2(i)\}$
so that $\{\pi_{h_i}(i)\}_{i=1}^{n-1}=\{1,2,\ldots, n-1\}$. So
$$\Pi_{i=1}^{n-1}\frac{|x_{\pi_{h_i}(i)}|^2}{|x_i|^2}=1,$$ and thus,
by  Lemma 2.2 with $k=1$ and $s=n-1$, we get
$$\begin{array}{rl} &
\sum_{i=1}^{n}{\frac{|x_i|^2}{(n-2)|x_i|^2+|x_{\pi_1(i)}|^2+|x_{\pi_2(i)}|^2}}
=
\sum_{i=1}^{n-1}{\frac{1}{(n-2)+\frac{|x_{\pi_1(i)}|^2}{|x_i|^2}+\frac{|x_{\pi_2(i)}|^2}{|x_i|^2}}}\\
\leq &
\sum_{i=1}^{n-1}{\frac{1}{((n-1)-1)+\frac{|x_{\pi_{h_i}(i)}|^2}{|x_i|^2}}}\leq
\frac{n-1}{n-1}=1.
\end{array} \eqno(3.4)$$
 If $r\leq n-2$, then we always have
$$\begin{array}{rl} &
\sum_{i=1}^{n}{\frac{|x_i|^2}{(n-2)|x_i|^2+|x_{\pi_1(i)}|^2+|x_{\pi_2(i)}|^2}}
=\sum_{i=1}^{r}{\frac{1}{(n-2)+\frac{|x_{\pi_1(i)}|^2}{|x_i|^2}+\frac{|x_{\pi_2(i)}|^2}{|x_i|^2}}}\leq
\frac{r}{n-2}\leq 1.
\end{array} \eqno(3.5)$$

The inequalities in Eqs.(3.4) and (3.5) ensure us that $f$ is upper
bounded by 1 on the boundary of its domain, and hence Eq.(3.2) is
true
 for any unit vectors. Applying Lemma 3.1, $\Phi_D$ is a
positive linear map for the case $n\geq 4$.

{\bf Case 2.} $n=3$.

Finally, assume that $n=3$. If one of $\pi_1$ and $\pi_2$ is the
identity, then the question  reduce to the case of $D$-type maps
generalized by one permutation, and hence the corresponding $D$-type
map is positive by \cite{HLPQS}. So we may assume that both $\pi_1$
and $\pi_2$ are not the identity. It is easily checked that we have
four possible cases so that $\{\pi_1,\pi_2\}$ has the property (C):
$$\begin{array}{llll} \{(2,3,1),(3,1,2)\}; & \{(3,2,1),(2,1,3)\};
&\{(3,2,1),(1,3,2)\};& \{(1,3,2),(2,1,3)\}. \end{array}
$$
The case $\{(2,3,1),(3,1,2)\}$ is implied by the result in
\cite{SY}. By Lemma 3.1, the last three cases induce to prove the
following inequality
$$\frac{1}{1+u+v}+\frac{u}{2u+1}+\frac{v}{2v+1}\leq 1
$$
for any $u,v>0$. This is true because
$$\begin{array}{rl} &(2u+1)(2v+1)+u(1+u+v)(2v+1)+v(1+u+v)(2u+1) \\ \leq
&(1+u+v)(2u+1)(2v+1) \end{array}
$$
by $2uv\leq u^2+v^2$. So the theorem is also true for the case
$n=3$, finishing the proof. \hfill$\Box$

It is clear that  the less $n$ is the easier to check the property
(C) of $\{\pi_1,\pi_2
\}$. This motivates us to decompose the
permutations into small ones.

For a nonempty   subset $F$ of $\{1,2,\ldots, n\}$, if $F$ is a
common invariant subset of $\pi_s$s, i.e., if $\pi_h(F)=F$ holds
for all $h=1,2,\ldots k$, we say that $F$ is an invariant subsets
of $\{\pi_1,\pi_2,\ldots,\pi_k\}$. Obvious, there exist disjoint
minimal invariant subsets $F_1,F_2,\ldots, F_l$ of
$\{\pi_1,\pi_2,\ldots,\pi_k\}$ such that $\sum _{s=1}^l \# F_s=n$
(i.e., $\cup_{s=1}^l F_s=\{1,2,\ldots,n\}$). We say
$\{F_1,F_2,\ldots, F_l\}$ is the complete set of minimal invariant
subsets of $\{\pi_1,\ldots,\pi_k\}$. Thus one can reduce the set
$\{\pi_1,\pi_2,\ldots,\pi_k\}$ of permutations into $l$ sets
$\{\pi_{1s},\pi_{2s},\ldots,\pi_{ks}\}_{s=1}^l$ of small ones,
where $\pi_{hs}=\pi_h|_{F_s}$. It is easily checked that
 $\{\pi_1,\pi_2\}$ has the property (C) if and only if each pair $\{\pi_{1s},\pi_{2s}\}$ of
 the
sub-permutations  has the property (C).

 The following is a   version of Theorem 3.3.

{\bf Theorem 3.4.} {\it  Let $\{F_s \}_{s=1}^l$ be the complete set
of minimal invariant subsets of a pair $\{\pi_1,\pi_2 \}$ of
permutations of $\{1,2,\ldots,n\}$. Let $\pi_{is}=\pi_i|_{F_s}$. If
$\{\pi_{1s},\pi_{2s} \}$ has Property (C) for every $s=1,2,\ldots,
l$, then the $D$-type map $\Phi_D$ defined as in Theorem 3.3 is
positive.}

\if {\bf Proof.} For any scalars $x_1,x_2,\ldots, x_n$ with
$|x_1|^2+\cdots+|x_n|^2=1$, let $N=\{i: x_i\not=0\}$ and $F_s^\prime
=F_s\cap N$. Let $n_s=\#F_s$, $r_s=\#F_s^\prime$. Since
$\{\pi_{1s},\pi_{2s}\}$ has Property (C),
 by
 Lemma 2.2, Lemma 3.1 and the proof of Theorem 3.3,  if $r_s=n_s$ or $n_s-1$, we have
$$\begin{array}{rl}
  & \sum_{i\in F_s}{\frac{|x_i|^2}{(n-2)|x_i|^2+|x_{\pi_1(i)}|^2+|x_{\pi_2(i)}|^2}}\\
 =& \sum_{i\in F_s^\prime}{\frac{|x_i|^2}{(n-2 )|x_i|^2+|x_{\pi_1(i)}|^2+|x_{\pi_2(i)}|^2}}\\
 =& \sum_{i\in F_s^\prime}{\frac{1}{(n-2)
 +\frac{|x_{\pi_1(i)}|^2}{|x_i|^2}+\frac{|x_{\pi_2(i)}|^2}{|x_i|^2}
 }}\\
 \leq&  \frac{r_s}{n-n_s+r_s}\leq\frac{n_s}{n} ;
\end{array}
 $$
 if $r_s \leq n_s-2$, we have
 $$\begin{array}{rl}
  & \sum_{i\in F_s}{\frac{|x_i|^2}{(n-2)|x_i|^2+|x_{\pi_1(i)}|^2+|x_{\pi_2(i)}|^2}}\\
 =& \sum_{i\in F_s^\prime}{\frac{1}{(n-2) +\frac{|x_{\pi_1(i)}|^2}{|x_i|^2}+\frac{|x_{\pi_2(i)}|^2}{|x_i|^2}|x_i|^2
}}\\
 \leq&  \frac{r_s}{n-2}\leq\frac{n_s-2}{n-2} \leq\frac{n_s}{n}.
\end{array}
 $$
It follows that
 $$\begin{array}{rl} &
 \sum_{i=1}^{n}{\frac{|x_i|^2}{(n-2)|x_i|^2+|x_{\pi_1(i)}|^2+|x_{\pi_2(i)}|^2}}\\
 =&\sum_{s=1}^l\sum_{i\in
 F_s}{\frac{|x_i|^2}{(n-2)|x_i|^2+|x_{\pi_1(i)}|^2+|x_{\pi_2(i)}|^2}}\\
 \leq&
 \sum_{s=1}^l\frac{n_s}{n}=1,
\end{array}
 $$
and hence, $\Phi_D$ is positive.\hfill$\Box$ \fi

 Theorem 3.3 and Theorem 3.4 give a possible way to construct new positive
 maps but it is not very clear how to check whether or not  a given pair of permutations has the
 property (C). To get an insight of this,
in the following, we give a characterization of a pair of
permutations to have property (C), which will be used to construct
new positive maps in the next section.  Note that
$\{\pi_1,\pi_2\}$ has property (C) if and only if each pair
$\{\pi_{1s},\pi_{2s}\}$ on the common invariant subset $F_s$ has
property (C). So, to check whether or not $\{\pi_1,\pi_2\}$ has
property (C), we may assume that $\pi_1$ and $\pi_2$ has no common
proper invariant subsets. It is clear that, if $n=1$,
$\{\pi_1,\pi_2\}$ always has property (C); if $n=2$,
$\{\pi_1,\pi_2\}$ has property (C) if and only if one of
$\pi_1,\pi_2$ is the identity and other is
$(1,2)\rightarrow(2,1)$.

{\bf Proposition 3.5.} {\it Let $\pi_1,\pi_2$ be two permutations
of $(1,2,\ldots,n)$ with $n\geq 2$ having no proper common
invariant subsets. Then $\{\pi_1,\pi_2\}$ has the property (C) if
and only if the following conditions   are satisfied:}

(1) {\it For any distinguished $i,j$, $\pi_1(i)\not=\pi_2(i)$ and
$\{\pi_1(i),\pi_2(i)\}\not=\{\pi_1(j),\pi_2(j)\}$; }

(2) {\it For any $i$ and  $j_1,j_2$ with $\pi_1(j_2)=\pi_2(j_1)=i$,
if distinct $j_3,\ldots,j_m \not\in\{i,j_1,j_2\}$ satisfy that
$\pi_2(j_3)=\pi_1(j_1)$, $\pi_2(j_4)=\pi_1(j_3),\ldots,
\pi_2(j_m)=\pi_1(j_{m-1})$, then $\pi_1(j_m)\not=\pi_2(j_2)$.}

{\bf Proof.} Assume that $\{\pi_1,\pi_2\}$ satisfy the conditions
(1)-(2). For any $i$,  we have to show that we can choose one
element in $\pi_{h_j}(j)\in\{ \pi_1(j),\pi_2(j)\}$ for each
$j\not=i$ so that $\{\pi_{h_j}(j), j\not=i\}=\{1,2,\ldots,
i-1,i+1,\ldots, n\}$.

Case (i). $i\in\{\pi_1(i),\pi_2(i)\}$, say $\pi_1(i)=i$, then
obviously the choice $\{\pi_1(j): j\not=i\}=\{1,2,\ldots,
i-1,i+1,\ldots, n\}$.

Case (ii). $i\not\in\{\pi_1(i),\pi_2(i)\}$.

Let $j_1,j_2$ such that $\pi_1(j_2)=i=\pi_2(j_1)$. By the condition
(1), we must gave $\pi_1(j_1)\not= \pi_2(j_2)$. In fact, if
$\pi_1(j_1)=\pi_2(j_2)$, then
$\{\pi_1(j_1),\pi_2(j_1)\}=\{\pi_1(j_2),\pi_2(j_2)\}$, which
contradicts to the condition (1). If $\{\pi_1(j_1),\pi_2(j_2)\}
=\{\pi_1(i),\pi_2(i)\}$, then
$\{\pi_1(j_1),\pi_2(j_2)\}\cup\{\pi_1(j):
j\not\in\{i,j_1,j_2\}\}=\{1,\ldots, i-1,i+1,\ldots, n\}$ and we
finished the proof.

Assume that $\{\pi_1(j_1),\pi_2(j_2)\}\not=\{\pi_1(i),\pi_2(i)\}$.

If $\pi_1(j_1)=\pi_2(i)$ or $\pi_2(j_2)=\pi_1(i)$, saying
$\pi_2(j_2)=\pi_1(i)$, then we have $\{\pi_2(j_2)\}\cup\{\pi_1(j):
j\not\in\{i,j_2\} \}=\{\pi_1(j):j\not=j_2\}= \{1,2,\ldots,
i-1,i+1,\ldots, n\}$, and then the proof is finished.

Thus we may in the sequel assume that
$\{\pi_1(j_1),\pi_2(j_2)\}\cap\{\pi_1(i),\pi_2(i)\}=\emptyset$. Take
$j_3$ so that $\pi_2(j_3)=\pi_1(j_1)$. As $\pi_1(j_1)
\not=\pi_2(j_2)$, we have $j_3\not=j_2$.  Also
$\pi_2(j_3)=\pi_1(j_1)\not=\pi_2(i)$ and
$\pi_2(j_3)=\pi_1(j_1)\not=\pi_2(j_1)=i $ ensures that
$j_3\not\in\{i,j_1\}$. So have $j_3\not \in\{i,j_1,j_2\}$ and by the
condition (3), we have $\pi_1(j_3)\not=\pi_2(j_2)$. Thus we get a
set $\{\pi_1(j_1),\pi_2(j_2),\pi_1(j_3)\}$ of distinct elements. If
$\pi_1(j_3)=\pi_2(i),$ then
$\{\pi_1(j_1),\pi_1(j_3)\}\cup\{\pi_2(j):j\not
\in\{i,j_1,j_3\}=\{\pi_2(j):j\not=j_1\}= \{1,2,\ldots,
i-1,i+1,\ldots, n\}$ and we finish the proof. Suppose that
$\pi_1(j_3)\not=\pi_2(i)$ and take $j_4$ so that
$\pi_2(j_4)=\pi_1(j_3)$. Then $j_4\not=i$ and $j_4\not=j_3$ by the
previous proof. In fact, because
$\pi_2(j_3)=\pi_1(j_1)\not=\pi_1(j_3)$, so
$\pi_2(j_3)\not=\pi_1(j_3)$, and $j_4\not=j_3$. As
$\pi_2(j_4)=\pi_1(j_3)\not=\pi_2(j_2)$, thus we have $j_4\not=j_2$.
Also $\pi_2(j_4)=\pi_1(j_3)\not=\pi_1(j_2)=\pi_2(j_1)=i$ implies
that $j_4\not=j_1$. So $j_4\not\in\{i,j_1,j_2,j_3\}$  and by the
condition (3), $\pi_1(j_4)\not=\pi_2(j_2)$. Therefore, we have
$\pi_1(j_4)\not\in\{\pi_1(j_1),\pi_2(j_2),\pi_1(j_3)\}$. Again, if
$\pi_1(j_4)=\pi_2(i),$ then
$\{\pi_1(j_1),\pi_1(j_3),\pi_1(j_4)\}\cup\{\pi_2(j):j\not
\in\{i,j_1,j_3,j_4\}=\{\pi_2(j):j\not=j_1\}=
\{1,2,\ldots,i-1,i+1,\ldots, n\}$, and the proof is finished. If
$\pi_1(j_4)\not=\pi_2(i)$, then, by the condition (3),
$\pi_1(j_4)\not=\pi_2(j_2)$ and $\pi_1(j_4)\not=\pi_2(j_1)=i$. It
follows that we can take $j_5\not\in\{i,j_1,j_2,j_3,j_4\}$ such that
$\pi_2(j_5)=\pi_1(j_4)$, and
$\pi_1(j_5)\not\in\{\pi_1(j_1),\pi_2(j_2),\pi_1(j_3),\pi_1(j_4)\}$.
We continue the above process until getting $j_m$ such that
$\pi_2(j_m)=\pi_1(j_{m-1})$ and $\pi_1(j_m)=\pi_2(i)$. Then we get
$\{\pi_1(j_1),\pi_1(j_3),\ldots,\pi_1(j_m)\}\cup\{\pi_2(j):j\not\in\{i,j_1,j_3,\ldots,j_m\}\}
=\{\pi_2(j):j\not=j_1\}=\{1,2,\ldots,i-1,i+1,\ldots, n\}$.
 Hence, the conditions (1) and (2) imply
$\{\pi_1,\pi_2\}$ has Property (C).

By checking the arguments above, one sees that, if any one of the
conditions (1) and (3) is broken, then $\{\pi_1,\pi_2\}$ can not
have Property (C). Therefore, the conditions are also necessary.
\hfill$\Box$

By the symmetry of $\pi_1$ and $\pi_2$, the condition (2) in
Proposition 3.5 may be replaced by the following

(2$^\prime$) {\it For any $i$ and  $j_1,j_2$ with
$\pi_1(j_2)=\pi_2(j_1)=i$, if distinct $j_3,\ldots,j_m
\not\in\{i,j_1,j_2\}$ satisfy that $\pi_1(j_3)=\pi_2(j_2)$,
$\pi_1(j_4)=\pi_2(j_3),\ldots, \pi_1(j_m)=\pi_2(j_{m-1})$, then
$\pi_2(j_m)\not=\pi_1(j_1)$.}\vspace{3mm}

{\bf Corollary 3.6.} {\it Let $\pi_1,\pi_2$ be permutations of
$(1,2,\ldots, n)$ and let $D=(n-2)I_n+P_{\pi_1}+P_{\pi_2}$.   Then
the $D$-type map $\Phi_D$ is positive if   for any minimal invariant
subset $F$ of $\{\pi_1,\pi_2\}$ with $\# F\geq 2$, $\{\pi_1|_F,
\pi_2|_F\}$ satisfies the conditions (1) and (2) in Proposition
3.5.}

{\bf Proof.} The corollary is an immediate consequence of
Proposition 3.5 and  Theorem 3.4. \hfill$\Box$

Before going to next section, we give a simple example of how to
using the results in this section to construct new positive linear
maps.

{\bf Example 3.7.}  The map $\Phi  :M_5\rightarrow M_5$ defined by
$$
(a_{ij})\mapsto\left(\begin{array}{ccccc} 3a_{11}+a_{55} & -a_{12} &-a_{13} & -a_{14}&-a_{15}\\
-a_{21} & 2a_{22}+a_{11}+a_{44}& -a_{23}& -a_{24} &-a_{25} \\
-a_{31} & -a_{32} & 3a_{33}+a_{22} & -a_{34} &-a_{35} \\
-a_{41} & -a_{42} & -a_{43}& 2a_{44}+a_{33}+a_{55}   &-a_{45} \\
-a_{51} &  -a_{52}& -a_{53}& -a_{54} & 2a_{55}+a_{44}
\end{array}\right)
$$
is positive. In fact, $\Phi$ is a $D$-type map  with
$D=3I_5+P_{\pi_1}+P_{\pi_2}$, where $\pi_1$ and $\pi_2$ are
permutations of $(1,2,3,4,5)$ determined respectively by
$(2,3,4,5,1) $ and $(1,5,3,2,4)$. It is easily checked by Definition
3.2 or Proposition 3.5 that $\{\pi_1,\pi_2\}$ has the property (C).
Hence the positivity of $\Phi$ follows from Theorem 3.3. We remark
here that, this example is different from those treated in the next
section.

\section{Constructing   positive $D$-type maps from cyclic permutations}

 The class of positive linear maps may detected by Theorem 3.3 is
quite large, which contains many known positive maps of $D$-type
such as that induced by just one permutation in \cite{HLPQS} and
that in \cite{SY} with $k=2$. Applying the results, especially
Theorem 3.3, Proposition 3.5 in the previous section, we construct
in this section some new classes of $D$-type positive maps induced
by two  powers  of the cyclic permutation.

In the sequel, for any integer $n\geq 3$, we denote $\pi$ the cyclic
permutation of $(1,2\ldots, n)$ defined by $\pi(i)=i+1$ mod $n$. Let
$1\leq p<q\leq n$ be two integers. Consider the $D$-type map induced
by $\pi^p,\pi^q$ with $D=(n-2)I_n+P_{\pi^p}+P_{\pi^q}$. We shall
give an easily handled criterion for $\Phi_{n,p,q}=\Phi_D$ to be
positive.
 Note that $q=n$ implies
$\pi^q={\rm id}$ and
$D=(n-2)I_n+P_{\pi^p}+P_{\pi^n}=(n-1)I_n+P_{\pi^p}$. So, $\Phi_D$ is
positive by \cite{HLPQS,QH} whenever $q=n$ and it is also clear that
$\{\pi^p,{\rm id}\}$ has the property (C). Thus we may assume $q<n$
in the following lemma.

{\bf Lemma 4.1.} {\it Let $\pi$ be the permutation of
$(1,2,\ldots,n)$ defined by $\pi(i)=i+1$ (mod $n$). For any $1\leq
p<q< n$, $\{\pi^p,\pi^q\}$ has the property (C) if and only if one
of the following conditions   holds.}

(1) {\it $q-p=1$.}

(2) {\it $1<q-p<n-1$ and  if there are   relatively prime positive
integers $k,m$ with $m<n$ and $k<q-p$ such that $m(q-p)= {kn} $,
then $p=n-d(q-p)$ for some integer $1\leq d\leq m-1$. }

{\bf Proof.} If $q-p=1$, then it is clear that $\{\pi^p,\pi^q\}$ has
property (C).

So, to prove the lemma, it suffices to show that, for any $p,q$ with
$1<q-p<n-1$ and $1\leq p<q\leq n-1$, $\{\pi^p,\pi^q\}$ does not have
property (C) if and only if there exist relatively prime integers
 $k,m$ with $0<k<q-p$ and $0<m<n$ such that $m(q-p)=kn$ and
$p\not=n-d(q-p)$ for any integer $d$ with $2\leq d\leq m-1$. By
Proposition 3.5, it suffices to check the following two claims.

{\bf Claim 1.}   $\{\pi_1=\pi^p,\pi_2=\pi^q\}$ does not satisfy the
condition (1)  of Proposition 3.5 if and only if   $2(q-p)=n$ and
$p\not=n-(q-p)$.

In fact, $\{\pi_1,\pi_2\}$ does not meet the condition (1) of
Proposition 3.5 if and only if $$i+p\ {\rm mod}\
n=\pi_1(i)=\pi_2(j)=j+q\ {\rm mod}\ n$$ and
$$i+q\ {\rm mod}\ n=\pi_2(i)=\pi_1(j)=j+p\ {\rm mod}\ n$$ for some
distinct $i,j$. This happens if and only if $2(q-p)=0$ mod $n$, and
in turn, if and only if $n=2(q-p)$ as $q\leq n $. It is clear that
$p$ can not be $n-(q-p)$, because this would imply that $q=n$.

{\bf Claim 2.}  $\{\pi_1=\pi^p,\pi_2=\pi^q\}$ does not satisfy the
condition (2)   of Proposition 3.5 if and only if   $m(q-p) =kn$
 for some  relatively prime positive integers
$k,m$ with $1\leq k<q-p$, $2< m<n$, and $p\not=n-d(q-p) $ for any
$1< d< m-1$.

$\{\pi_1=\pi^p,\pi_2=\pi^q\}$ does not satisfy the condition (2) of
Proposition 3.5 if and only if for some $i$, there exist distinct
$j_1, \ldots,j_m\not\in\{i\}$ such that $\pi_1(j_2)=\pi_2(j_1)=i$,
$\pi_2(j_3)=\pi_1(j_1)$, $\pi_2(j_4)=\pi_1(j_3),\ldots,
\pi_2(j_m)=\pi_1(j_{m-1})$ and $\pi_1(j_m)=\pi_2(j_2)$, and in turn,
if and only if for some $i$ there exist distinct $j_1,
\ldots,j_m\not\in\{i\}$ such that
$$
\begin{array}{rl}  j_1+q=&j_2+p\ {\rm mod }\ n,\\
 j_3+q=&j_1+p\ {\rm mod }\ n,\\
 j_4+q=&j_3+p\ {\rm mod }\ n,\\
\vdots & \\
 j_m+q=&j_{m-1}+p\ {\rm mod }\ n,\\
 j_2+q=&j_m+p\ {\rm mod }\ n.
\end{array} \eqno(4.1)
$$
Summing up all equations in Eq.(4.1) we obtain that
$$m(q-p)=0\ \ {\rm mod}\ n.$$ Note that $2\leq m<n$;
 thus   there exists positive integer  $k<q-p$ such
that
$$ m(q-p)=kn.
\eqno(4.2)
$$
Moreover, $j_h=j_1-(q-p)(h-2)=j_2-(q-p)(h-1)$ mod $n$ for
$h=3,4,\ldots, m$. Obviously, $j_1,j_2\not\in\{i\}$. While
$j_3,\ldots ,j_m\not\in\{i\}$ implies that $j_h\not=j_1+q=j_2+p$ mod
$n$ for any $h=3,4,\ldots, m$. It is clear that $j_h=i=j_1+q=j_2+p$
mod $n$ for some $j_h=j_1-(q-p)(h-2)=j_2-(q-p)(h-1)$ mod $n$ if and
only if $j_1+q=j_1-(q-p)(h-2)$ and $j_2+p=j_2-(q-p)(h-1)$, and in
turn, if and only if $p=n-(q-p)(h-1)=n-d(q-p)$ for some $2\leq d\leq
m-1$.  So, $j_1,\ldots ,j_m\not\in\{i\}$ implies that
$p\not=n-d(q-p) $ for any $2\leq d<m$. Thus we proved that
$\{\pi^p,\pi^q\}$ does not satisfies Condition (2) of Proposition
3.6 implies that $m(q-p)=kn$ for some relatively prime positive
integers $m, k$ with $1\leq k<q-p$, $1\leq m<n$, and
$$p\not=n-d(q-p)  \quad{\rm for\ every }\ 2\leq d\leq m-1.
\eqno(4.3)$$

 Conversely,
assume that Eq.(4.2) and Eq.(4.3) hold.   Take any $j_1,j_2$ so that
$j_2-j_1=q-p$ mod $n$. Then $j_2=j_1+(q-p)$ mod $n$. If $m=2$, we
must have $2(q-p)=n$ and hence $j_2+(q-p)=j_1+2(q-p)=j_1$ mod $n$,
which gives $j_2+q=j_1+p$ mod $n$. If $m\geq 3$, for
     $3\leq h\leq m$, let
$$ j_h=j_1-(q-p)(h-2)=j_2-(q-p)(h-1) \ \ {\rm
mod}\ n.
 \eqno(4.4)
$$
Then $j_2=j_m+(q-p)(m-1)$ mod $n$ and hence $j_2+q-p=j_m$ mod $n$ as
$(q-p)m=kn$. This ensures that $j_2+q=j_m+p$ mod $n$. We claim that
$j_h\not=j_s$ whenever $t\not=s$. In fact, it is clear that
$j_1\not=j_2$; for   $3\leq h,s\leq m$,
$$j_h=j_s\Leftrightarrow
j_1-(q-p)(h-2)=j_1-(q-p)(s-2)\Leftrightarrow h=s.$$ Moreover, the
Eqs.(4.3) and (4.4) imply that $j_1,\ldots, j_m\not\in\{i\}$. Hence
Eq.(4.2) and Eq.(4.3) imply that $\{\pi^p,\pi^q\}$ does not
satisfies the condition (2) of Proposition 3.5.

Now,  by Proposition 3.5, Claims 1-2 ensure that the lemma holds,
 completing the proof.\hfill$\Box$

Using Lemma 4.1, the following criterion of positivity of
$\Phi_{n,p,q}$ is immediate, which is very easily applied.

{\bf Theorem 4.2.} {\it Let $n\geq 3$ and $\pi$ be the permutation
defined by $\pi(i)=i+1$ mod $n$. For any integers $1\leq p< q\leq
n$, let $D=(n-2)I_n +P_{\pi^p}+P_{\pi^q}$. Then the $D$-type map
$\Phi_{n,p,q}=\Phi_D: M_n\rightarrow M_n$ of the form Eq.(1.1) is
positive if one of the following   conditions   holds.}

(1) {\it $q-p=1$ or $q=n$.}

(2) {\it $q<n$, $1<q-p<n-1$ and if there are   relatively prime
positive integers $k,m$ with $m<n$ and $k<q-p$ such that $m(q-p)=
{kn} $, then $p=n-d(q-p)$ for some integer $1\leq d\leq m-1$. }

We remark  that in general the condition $q-p=\frac{kn}{m}$ does not
imply that $q-p$ is a factor of $n$. For example, let $n=8$,
$q-p=6$; then $m=3$ and $k=4$. If $p=8-6=2,q=8$, then
$\{\pi^2,\pi^8\}$ has property (C). But, $\{\pi^1,\pi^7\}$ does not
posses property (C).

The following corollary is immediate.

{\bf Corollary 4.3.} {\it The $D$-type map $\Phi_{n,p,q}:
M_n\rightarrow M_n$ in Theorem 4.2 is positive if any one of the
following conditions  holds.}

(1) {\it $q-p=1$ or $q=n$.}

(2) {\it $n$ is prime.}

(3) {\it There are no relatively prime $m,k$ with $1\leq m<n$ and
$1\leq k<q-p$ so that $m(q-p)=kn$. }

(4) {\it   $q-p$ is prime and is not a factor of $n$. Particularly,
if $q-p=2$ and $n$ is odd.}

(5) {\it $q-p$ is a prime factor of  $n$ and $p=d(q-p)$ for some
$1\leq d\leq \frac{n}{q-p}-2$.}

{\bf Proof.} The implication (1) $\Rightarrow \Phi_D$ is positive is
clear   by Theorem 4.2.

If (2) holds, that is, if   $n$ is prime, then  there are no
relative prime positive integers $m<n,k<q-p$ such that $m(q-p)=kn$.
Hence,  for any $1\leq p<q \leq n$, $\{\pi^p,\pi^q\}$ has property
(C).

(3) entails that  $\{\pi^p,\pi^q\}$ meets the condition (2) in
Theorem 4.2 and hence   $\Phi_{n,p,q}$ is positive.

If (4) holds, that is,  if $q-p$ is prime and is not a factor of
$n$, then $n=\frac{m(q-p)}{k}$ implies that $m=kr$ for some integer
$r$ and $n=r(q-p)$, which is  impossible. So $\{\pi^p,\pi^q\}$ meets
the condition (2) of Theorem 4.2 and thus $\Phi_{n,p,q}$ is
positive.

The condition (5) implies that $m_1(q-p)=kn$ and the greatest common
divisor of $m_1$ and $k$ is 1 if and only if $m_1=m$ and $k=1$ as
$q-p$ is prime. As $p=d(q-p)=n-(m-d)(q-p)$, $\{\pi^p,\pi^q\}$ has
property (C) by Lemma 4.1. \hfill$\Box$

Since, in quantum information theory, an $N$-qbit quantum system
corresponds to a complex Hilbert space of dimension $2^N$, the case
$n=2^N$ is of special importance.

\textbf{Corollary 4.4.}\ \ For $n=2^N$ with $N\geq 2$ and $1\leq
p<q<n$,   the $D$-type map $\Phi_{{2^N},p,q}: M_n\rightarrow M_n$ in
Theorem 4.2 is positive if one of the following holds:

(1) $q-p$ is odd.

(2) $q-p=2^b$ with $1\leq b\leq N-1$ and $p=d2^b$ for some $1\leq
d\leq 2^{N-b}-1$.

(3) $q-p=2^br$ with $r$ odd and $p=2^b(2^{N-b}-dr)$ for some $1\leq
d\leq \frac{2^{N-b}-1}{r}$.

In the following we present some simple examples to illustrate how
to use the results in this section.

\textbf{Example 4.5.}\ \ For $n=5$ or $7$ and any $1\leq p<q\leq n$,
 the $D$-type maps $\Phi_{5,p,q}: M_5\rightarrow M_5$ and $\Phi_{7,p,q}: M_5\rightarrow M_5$
are always positive.

Observe that $\Phi_{5,1,3}$   is of  the form
$$ \left(\begin{array}{ccccc} a_{11}&a_{12}&a_{13}&a_{14}&a_{15}\\
a_{21}&a_{22}&a_{23}&a_{24}&a_{25}\\
a_{31}&a_{32}&a_{33}&a_{34}&a_{35}\\
a_{41}&a_{42}&a_{43}&a_{44}&a_{45}\\
a_{51}&a_{52}&a_{53}&a_{54}&a_{55}
\end{array}\right)\mapsto$$
$$ \small \left(\begin{array}{ccccc}2a_{11}+a_{22}+a_{44}&-a_{12}&-a_{13}&-a_{14}&-a_{15}\\
-a_{21}&2a_{22}+a_{33}+a_{55}&-a_{23}&-a_{24}&-a_{25}\\
-a_{31}&-a_{32}&2a_{33}+a_{44}+a_{11}&\-a_{34}&-a_{35}\\
-a_{41}&-a_{42}&-a_{43}&2a_{44}+a_{55}+a_{22}&-a_{45}\\
-a_{51}&-a_{52}&-a_{53}&a_{54}&2a_{55}+a_{11}+a_{33}
\end{array}\right), $$
which is positive by Corollary 4.3.

It is also obvious that when $n=4$,  $\Phi_{4,p,q}$ is   positive if
$q-p\not=2$; when $n=6$, $\Phi_{6,p,q}$ is positive if
$q-p\not\in\{2,3,4\}$.

{\bf Example  4.6.} For $n=8=2^3$, and any $1\leq p<q\leq 8$,  the
$D$-type map $\Phi_{8,p,q}: M_8\rightarrow M_8$  is positive if one
of the following is true:

(i) $q-p\in\{1,3,5,7\}$.

(ii) $q-p=2$, $p\in\{2, 4,6\}$.

(iii) $q-p=4$, $p=4$.

(iv) $q-p=6$, $p=2$.

{\bf Example  4.7.} For $n=16=2^4$, and any $1\leq p<q\leq 16$, the
$D$-type map $\Phi_{16,p,q}: M_{16}\rightarrow M_{16}$  is positive
if one of the following is true:

(i) $q-p\in\{1,3,5,7,9,11,13,15\}$.

(ii) $q-p=2$ and $p\in\{2,4,6,8,10,12,14\}$.

(iii) $q-p=4$ and $p\in\{4, 8,12\}$.

(iv) $q-p=6$ and $p\in\{4,10\}$.

(v) $q-p=8$ and $p=8$.

(vi) $q-p=10$ and $p=6$.

(vii) $q-p=12$ and $p=4$.

(viii) $q-p=14$ and $p=2$.

{\bf Remark 4.8.} The condition that $\{\pi^p,\pi^q\}$ has property
(C)  in Theorem 4.2 is sufficient for the associated $D$-type map
$\Phi_{n,p,q}$ to be positive. The condition is not necessary, and
thus, the condition in Theorem 3.3 is not necessary for $\Phi_D$ to
be positive. For example, let us consider the case of $n=4$. Then,
$\{\pi^p,\pi^q\}$ does not have property (C) if and only if $p=1,
q=3$. By Lemma 3.1, the $D$-type map $\Phi_{4,1,3}=\Phi_D$ with
$D=2I_4+P_{\pi}+P_{\pi^3}$ is positive if and only if
$$f(x_1,x_2,x_3,x_4)=\frac{x_1}{2x_1+x_2+x_4}+\frac{x_2}{2x_2+x_3+x_1}+\frac{x_3}{2x_3+x_4+x_2}+\frac{x_4}{2x_4+x_1+x_3}\leq
1
$$
holds for all non-negative $x_1,\ldots, x_4\in{\mathbb R}$ with
$x_1+\cdots+x_4=1$. By Lemma 2.2, all extremum values of
$f(x_1,x_2,x_3,x_4)$ is $1$ because $s=M=2$. Consider the supremum
of $f$ on the boundary.  Assume that only one of $x_i$ is zero, say
$x_4=0$; then
$$f(x_1,x_2,x_3,0)=\frac{x_1}{2x_1+x_2 }+\frac{x_2}{2x_2+x_3+x_1}+\frac{x_3}{2x_3
+x_2}
$$
with $x_1,x_2,x_3$ nonzero. Consider the function
$$g(s,t)=\frac{1}{2+s+t}+\frac{1}{2+\frac{1}{s}
}+\frac{1}{2+\frac{1}{t} }=\frac{1}{2+s+t}+\frac{s}{2s+1
}+\frac{t}{2t+1},$$  where $s>0$ and $t>0$. As
$$\begin{array}{rl} &(2s+1)(2t+1)+s(s+t+2)(2t+1)+t(s+t+2)(2s+1)\\ =&
4s^2t+4st^2+14st+s^2+t^2+4s+4t+1,\end{array}
$$
$$
(2s+1)(2t+1)(s+t+2)=4s^2t+4st^2+12st+2s^2+2t^2+5s+5t+2
$$
and $2st<s^2+t^2+s+t+1$, it is easily checked that
$g(s,t)=1-\frac{(s-t)^2+s+t+1}{(2s+1)(2t+1)(s+t+2)}<1$. So we still
have $f(x_1,x_2,x_3,0)<1$. If there are more than one $x_i=0$, it is
clear that $f(x_1,x_2,x_3,x_4)<1$. Therefore, $\sup
f(x_1,x_2,x_3,x_4)=1$ on the region $x_1+\cdots +x_4=1$ and $\Phi_D$
is positive.

It is then interesting to ask

{\bf Question 4.9.} What is the necessary and sufficient condition
for $\Phi_{n,p,q}=\Phi_D: M_n\rightarrow M_n$ with
$D=(n-1)I_n+P_{\pi^p}+P_{\pi^q}$ to be positive? Where $\pi$ is the
permutation defined by $\pi(i)=i+1$ mod $n$ and $1\leq p<q< n$.


\end{document}